# Decreasing defect rate of test cases by designing and analysis for recursive modules of a program structure: Improvement in test cases

Muhammad Javed[1], Bashir Ahmad[1], Zaffar Abbas[1], Allah Nawaz[1], Muhammad Ali Abid[1], Ihsan Ullah[1]

[1]Institute of Computing and Information Technology Gomal University, D.I.Khan, Pakistan

*Abstract---**Designing and analysis of test cases is a challenging** tasks **for tester roles especially those who are related to test the structure of program. Recently, Programmers are showing valuable trend towards the implementation of recursive modules in a program structure. In testing phase of software development life cycle, test cases help the tester to test the structure and flow of program. The implementation of well designed test cases for a program leads to reduce the defect rate and efforts needed for corrective maintenance. In this paper, author proposed a strategy to design and analyze the test cases for a program structure of recursive modules. This strategy will definitely leads to validation of program structure besides reducing the defect rate and corrective maintenance efforts.***

Index Term—Test cases, Recursive module, Black-box, White-box, corrective maintenance, defect rate.

## I. INTRODUCTION

Testing phase of software development life cycle lead to the quality of software products and it depend on the strategies which are followed by tester role. The most commonly used methods of testing are black-box and white-box testing [1]. In black-box tester examine the fundamentals aspects of software; while in white-box tester examine the internal procedure detail of the system components such path testing and loop testing. During white-box testing test cases can be generated either manual or through automated tool to check the working of software. A test case is a set of conditions or variables which are included in the working of software[5,7]. The focus of this paper is to design and analyze the generation of test cases for recursive modules in programming language. Here author's proposed a strategy which helps to reduce the defect rate and corrective maintenance efforts.

## II. RECURSIVE MODULES

In programming language structure recursive modules are those routines which called itself during execution of program and they can consider as central idea of computer science [6, 8]. There are two factors which are relevant to recursive modules. First is the base case used to end the calling of recursive module and second is to break the current domain of data into sub domains and this will remain continue till base case satisfied [10,11]. Recursive modules are classified into linear, mutual, binary and N-Ary Types.

## III. WHITE BOX TESTING

White box testing is the process to test the implementation of a system. It consist of analysis of data flow, control flow, information flow , coding practice and exception handling within the system to ensure correct software behavior. White box testing can be perform by tester any time after coding but it will be good practice to do it with unit testing. White box testing is used with unit, integration and regression testing. In white box testing method tester role can perform the following activities[3].

- It defines the test strategy and activities.
- It develop new test plan on the base of selected strategy.
- It creates an environment for test case execution.
- It executes the test cases and prepared reports.

The main types of white box testing are static and dynamic analysis, branch coverage, security testing and mutation testing. Selection of skilled tester and bit of code to remove error are considered as important challenge in white box testing [9].

In software project the success of testing depend on the test cases used. To reduce the turn around time, defect rate and project duration it is important to design an effective set of test cases that enables detection of maximum number of errors [12].

## IV. FLOW GRAPH NOTATION (FGN)

In white-box testing Flow Graph Notation (FGN) is a used to represent the program control structure. It is just like flowchart and comprises on circle and edges. Each circle, called a flow graph node, represents one or more procedural statements and edges represent the flow of control. An edge must terminate at a node, even if the node does not represent any procedural statements. Areas bounded by edges and nodes are called regions. When counting regions, we include the area outside the graph as a region.

## V. ANALYSIS OF TEST CASES FOR RECURSIVE MODULES

To represents the analysis and design process of recursive modules an example in C++ language is taken as shown in Fig-1. In this example two recursive modules/functions are used named as "Factorial" and "SumofFact". Following steps are used to represent the working of C++ program shown in Fig-1.

```cpp
#include<iostream.h>
#include<stdio.h>
#include<conio.h>
int Factorial(int);
int SumofFact(int);
int sum=0,temp;
void main()
{
    clrscr();
    int number;
    cout<<"Enter Number to find factorial";
    cin>>number;
    cout<<"Factorial of number is = " << Factorial(number);
    cout<<"\nSum of factorials of all number from 1 to n"<<SumofFact(number);
    getch();
}
int Factorial(int n)
{
    if(n<1)
        return(1);
    else
        return(n*Factorial(n-1));
}
int SumofFact(int n)
{

    if(n>0)
     { sum=sum+Factorial(n);
       temp=SumofFact(n-1);}
    else
      return(sum);
}
```

**Fig-1. C++ Program including recursive modules/functions**

- Firstly a number is read in the main module/function of program.
- Secondly a recursive module named "Factorial" is called from main function to find the factorial of entered number. If number is 4 then result of "Factorial" function will be 24 i.e. 4! = 24.
- In third step another recursive module named "SumofFact" is called to add the sum of factorial of all numbers ranges from 1 to entered number. If number is 4 then result of this function will be 1!+2!+3!+4!=33

To analyze the complexity of program (shown in Fig-1) a Flow Graph Notation is drawn which is shown in fig-2. This FGN represent the all paths which can be used to analyze and design the test case for program. As there two factors, which are related with recursive module, first is the base condition which is applied to end the calling of recursive modules and second factor is relevant to division of domain of data for recursive module into sub domains. The complexity of recursive module calling can be analyzed with respect to two aspects.

1. Calling of a recursive module from any other module which is not recursive in nature.
2. Calling of a recursive module from any other module which is recursive in nature.

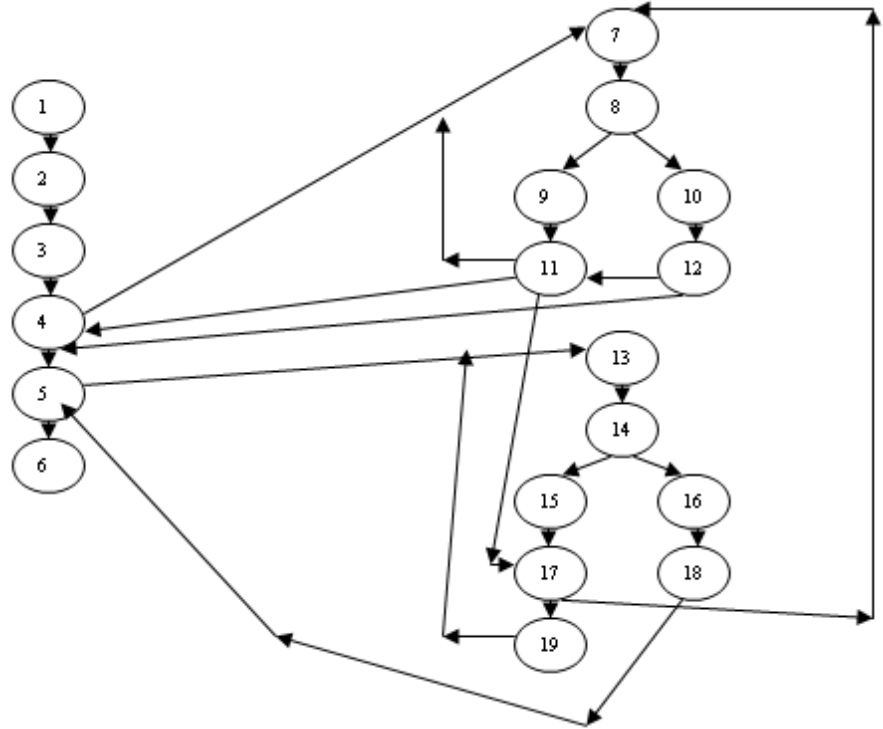

**Fig-2. Flow Graph Notation for C++ Program of recursive modules**

The complexity of program will be high for second aspect as compared to first. The program shown in Fig-1 represents the both aspect of calling the recursive module. The first aspect is represented through calling of "Factorial" recursive module/function and second is through calling of "SumofFact". In "SumofFact" recursive module/function "Fact" is again called. This process leads to increase the complexity of program.

*Analysis and Designing of Test cases and test data for the first Aspect:*
The first aspect shows the calling of recursive module/function from another function which is not recursive in nature. If we omit the "SumofFact" recursive function from program shown in Fig-1 and its calling from main module. Then there be will only two possible path to represent the execution of "Factorial" recursive module.

*Path-1.*
   1-2-3-4-7-8-10-12-5

*Path-2.*
   1-2-3-4-**7-8-9-11**-7-8-10-12-**11**-5

The first path represents the execution of statements of calling and called module in sequence. Which show the recursive module "Factorial" is called only one time from "main" module and it is not called by itself. The test case for this path will be n<1 and test data for this test case may be 0 or any negative number.

The second path represent that recursive module is called many time depend on the domain of data, this is shown in highlighted part of path i.e. 7-8-9-11. When the recursive module is called by itself last time then base condition will be executed which is shown in 7-8-10-12 part of second path. After that control will be transfer by recursive module to itself, this is shown in another highlighted part of second path i.e. 11. At the end control will be transfer back to the "main" calling module of recursive module. The test case for this path will be n>=1 and test data may be any positive value. If one recursive module is called many times from "main" calling module then same two paths will be used except the nodes of FGN will be increases. It is clear from this analysis that test case and test data will remain same whether you will called a recursive module one or more than one time.

*Analysis and Designing of Test cases and test data for second Aspect:*
The second aspect shows the calling of recursive module/function from another function which is recursive in nature. According to program of Fig-1 "SumofFact" is the calling module of "Factorial" recursive module and "SumofFact" itself is recursive module. To analyze the test cases for this aspect firstly omit the node 4 from FGN of "main" module. This will show that "Factorial" recursive module will not called from "main" module. There be will only two possible path to represent the execution of "SumofFact" and "Factorial" recursive modules.

*Path-1.*
   1-2-3-5-13-14-16-18-6

*Path-2.*
   1-2-3-5-**13-14-15-17-7-8-9-11-19**-13-14-16-18-6

The first path represents the execution of statements of calling (i.e. "main" module) and called module(i.e "SumofFact") in sequence. In this path execution of "Factorial" recursive module is not shown because here the base condition of "SumofFact" is executed and it not called itself. The test case for this path will be n<0 and test data may be 0 or any negative number. The second path represents the more than one time execution of "SumofFact" and "Factorial" modules. The part of second path i.e. 13-14-15-17-7-8-9-11-19 as a whole represents the recursive execution of both modules. In this part of second path i.e. 13-14-15-17 represents the execution of "SumofFact" and calling of "Factorial" recursive modules. Moreover, in this part of second path i.e. 7-8-9-11 represents the execution of "Factorial" recursive module and returning control back to the node 19 of FGN. This node also represents the calling of "SumofFact" recursive module i.e. same process will be repeated till the test case n<0 is satisfied. The test case for this path will be n>0 and test data may be any positive number. Moreover, from this analysis it is clear that first path eliminate the execution of base condition of "Factorial" recursive module, but it will not true for all cases. This is also illustrating here that during analysis and designing of test cases, some test cases can not show the execution of some part of a recursive module. So there is need to be more care during analysis and designing of test cases of recursive modules especially when a recursive module call another recursive module. If tester role will not care about it then it can leads to increase the defect rate and corrective maintenance efforts. Besides caring of tester role in analysis of recursive modules, it must care about the levels. If level to call one recursive module within another recursive module is increases then complexity of program will high and it will leads towards increases in defect rates.

## VI. CONCLUSION

During white-box testing process the use of FGN and deriving path are the basis steps to analyze and design the test cases and test data. In this paper authors adopt a strategy to analyze and design the test cases for recursive modules, which are considered as important paradigm in programming language. After analysis and designing process of test case authors known that some part of the recursive modules can not be implemented through test case which can increase the defect rate and corrective maintenance efforts.